\documentclass[twocolumn,preprintnumbers,longbibliography]{revtex4-2}

\usepackage{hyperref}
\usepackage{graphicx}
\usepackage{epsf}
\usepackage{amsmath,amssymb}
\usepackage{color}

\def\Sn#1#2#3{{}^{#1#2#3}\textrm{Sn}}
\def\SnA{{}^{A}\textrm{Sn}}

\usepackage{ulem}

\begin{document}
\title{Neutron $3s_{1/2}$ occupation change across the stable tin isotopes
investigated using isotopic analysis of proton scattering at 295~MeV}

\author{Yoshiko Kanada-En'yo}
\affiliation{Department of Physics, Kyoto University, Kyoto 606-8502, Japan}

\begin{abstract}
The isotopic systematics of Sn isotopes in the range $A=116$--124 were investigated
by combining the nuclear structure and reaction calculations for the analysis of
$\SnA(p,p)$ reactions at 295~MeV. The $\SnA(p,p)$ reactions were
calculated employing relativistic impulse approximation~(RIA) with
theoretical densities obtained for the Sn isotopes from
relativistic Hartree--Bogoliubov~(RHB) calculations with the DD-ME2 interaction and
nonrelativistic Skyrme Hartree--Fock--Bogoliubov~(SHFB) calculations with the SKM* interaction.
Calculation using the DD-ME2 density
reproduced the experimental data for $\Sn122(p,p)$ but overestimated the $\Sn116(p,p)$
and $\Sn118(p,p)$ cross sections at backward angles.
In the isotopic analysis of the cross section ratio $R(\sigma)$ of $\Sn122$ to $\Sn116$, a calculation using
the SKM* density reproduced the peak amplitudes of $R(\sigma)$ obtained from the experimental cross sections, whereas a calculation using the DD-ME2 density did not.
The ratio $R(\sigma)$ was found to be sensitive to the isotopic
change of the neutron $3s_{1/2}$ occupation through the isotopic difference of the surface neutron density around $r=4$--$5$~fm.
Isotopic analysis indicated
that a rapid increase of the $3s_{1/2}$ occupation
from $\Sn116$ to $\Sn122$ obtained by the DD-ME2 calculation is
unlikely.
This result derived from the 295 MeV $\SnA(p,p)$
cross section data \cite{Terashima:2008zza}
is consistent with the direct measurement of the neutron occupations
in Sn isotopes by neutron transfer and pick-up
reactions~\cite{Szwec:2021uwo}.

\end{abstract}

\maketitle

\section{Introduction}\label{sec:introdution}
Proton elastic scattering is a useful tool for determining not only the
neutron skin thickness of nuclei but also the detailed 
neutron density profile---in particular, the surface neutron
density---as has been performed for various nuclei~\cite{Ray:1978ws,Ray:1979qv,Hoffmann:1980kg,Terashima:2008zza,Zenihiro:2010zz,Zenihiro:2018rmz}.
In recent years, experiments on 295~MeV $(p,p)$ reactions \cite{Terashima:2008zza,Zenihiro:2010zz,Zenihiro:2018rmz} have been intensively performed
to extract the neutron density by reaction analysis using
the relativistic impulse approximation (RIA) with
density-dependent effective $NN$ interactions modified from
the original Murdock and Horowitz 
model~\cite{Horowitz:1985tw,Murdock:1986fs,RIAcode:1991}, 
called the ddMH model in this paper.
The RIA+ddMH model~\cite{Sakaguchi:1998zz} has 
successfully described the 295~MeV $(p,p)$ reactions of various target nuclei, including
Sn~\cite{Terashima:2008zza}, Pb~\cite{Zenihiro:2010zz}, and Ca~\cite{Zenihiro:2018rmz} isotopes.

$(p,p)$ cross sections are useful not only to
determine the neutron radii, but also to extract information about the structure properties
such as neutron single-particle properties and nuclear deformations, 
through their effects on the neutron density, which can be sensitively probed by the proton scattering.
In a previous paper~\cite{Kanada-Enyo:2021oee},  295~MeV $\textrm{Pb}(p,p)$ reactions were analyzed, 
and it was demonstrated 
that the cross sections were sensitive to the
single-particle occupation of the low-$\ell$ orbit in the major shell, 
which significantly contributes to the surface neutron
density. A detailed analysis was performed focusing on the
isotopic systematics of the nuclear structure properties and the reaction cross sections in a series of  Pb isotopes. Isotopic analysis
was demonstrated to be useful to reduce the systematic
uncertainty  in the data and the model ambiguities in the reaction analysis.

The present study aimed to extract the structure information on Sn isotopes
from the experimental data of $\SnA(p,p)$ cross sections at 295~MeV
for $A=116$--$124$~\cite{Terashima:2008zza}
by isotopic analysis combining  structure and reaction calculations.
The structure of the Sn isotopes was calculated  using both
relativistic Hartree--Bogoliubov~(RHB) and nonrelativistic Skyrme Hartree--Fock--Bogoliubov~(SHFB) calculations. Using the theoretical densities,
the Sn$(p,p)$ reactions were calculated
with the RIA+ddMH model
in the same way as in a previous study~\cite{Kanada-Enyo:2021oee}.
By comparing the calculated results and the experimental data,
the isotopic systematics of the structure
and reaction properties were investigated.
Particular attention was paid to the change of the $3s_{1/2}$ neutron occupation
from $\Sn116$ and $\Sn122$, which can be sensitively
probed by the isotopic systematics of the $(p,p)$ cross sections
through the effect on the surface neutron density.

This paper is organized as follows. The structure and reaction calculations
are explained in Section~\ref{sec:calculations}, and the calculated results
are presented in Section~\ref{sec:results}.
In Section~\ref{sec:analysis},
the isotopic analysis is performed.
Finally, a summary is provided in Section~\ref{sec:summary}.

\section{Calculations of nuclear structure and proton elastic scattering}\label{sec:calculations}
\subsection{Structure calculations}
The structure of even--even Sn isotopes with $A=114$--$124$
was calculated by employing the spherical RHB and SHFB calculations
using the computational DIRHB code~\cite{Niksic:2014dra} and HFBRAD code~\cite{Bennaceur:2005mx},
respectively.
In the RHB calculations, the
DD-ME2~\cite{Lalazissis:2005de} and DD-PC1~\cite{Niksic:2008vp} interactions were used,
which are denoted as me2 and pc1, respectively, in this paper.
In the SHFB calculations, the SKM*~\cite{Bartel:1982ed}
interaction with a mixed-type pairing force was used.
SHFB calculations with the SLy4~\cite{Chabanat:1997un} interaction were also performed;
however, the resulting densities of the Sn isotopes were
similar to those obtained using the SKM* interaction.

\subsection{Calculations of proton elastic scattering reactions}

The $\SnA(p,p)$ reactions at $E_p=295$~MeV were calculated using the RIA+ddMH model~\cite{Sakaguchi:1998zz},
in which real and imaginary nucleon-nucleus potentials were obtained by folding
the target density with the effective $NN$ interactions of the meson-exchange model.
The RIA+ddMH model was constructed by
introducing the density-dependent
$\sigma$- and $\omega$-meson masses and coupling constants for
the effective $NN$ interactions in the original
Murdock and Horowitz model~\cite{Horowitz:1985tw,Murdock:1986fs,RIAcode:1991}.
In the present calculation, 
the latest version of the parameterization of the RIA+ddMH model, which has been calibrated
to fit the $^{58}\textrm{Ni}(p,p)$ data at $295~$MeV~\cite{Zenihiro:2010zz}, was employed.

The RIA+ddMH calculation was performed using the theoretical densities of the Sn isotopes obtained by the RMF and SHFB calculations.
As structure inputs for calculating proton-nucleus potentials,
the theoretical neutron $(\rho_n(r))$ and proton $(\rho_p(r))$
densities of the target nuclei were used for the neutron and proton vector densities, while $0.96\rho_n(r)$
and $0.96\rho_p(r)$ were used for the neutron and proton scalar
densities, respectively, as performed in a previous study~\cite{Kanada-Enyo:2021oee}.
This prescription is the same as that performed in
the analysis of the $\textrm{Sn}(p,p)$ and $\textrm{Pb}(p,p)$ reactions
in Refs.~\cite{Terashima:2008zza,Zenihiro:2010zz}.

\section{Results}\label{sec:results}

\subsection{Radii and densities of the Sn isotopes}

\begin{figure}[!h]
\includegraphics[width=8 cm]{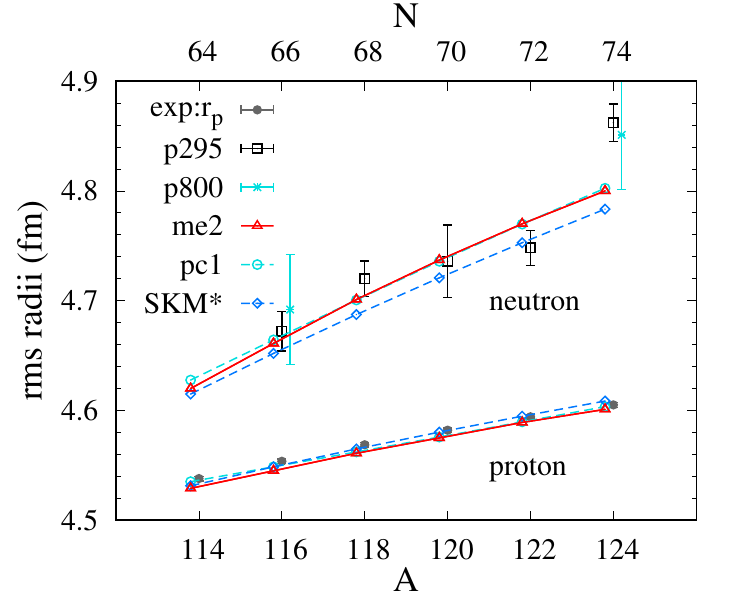}
\caption{
Rms neutron ($r_n$) and proton $r_p$ radii of Sn isotopes.
The theoretical values are the RHB (me2 and pc1) and
the SHFB (SKM*) calculations.
The experimental $r_n$ values are those determined 
by the $(p,p)$ reactions at both 295~MeV\cite{Terashima:2008zza} and 800~MeV~\cite{Ray:1979qv}
(p295 an dp800, respectively),
while the experimental $r_p$ values are obtained from the rms charge radii 
observed by isotope-shift measurements~\cite{Angeli:2013epw}.
\label{fig:rmsr}}
\end{figure}

The root-mean-square~(rms)  neutron $(r_n)$ and proton $(r_p)$ radii of the Sn isotopes
using the RHB (me2 and pc1) and SHFB (SKM*) calculations are presented in
Fig.~\ref{fig:rmsr}~(a), together with the experimental data.
In all calculations, the theoretical $r_p$ values reproduce
the experimental data quite well. For $r_n$, the theoretical values are in
reasonable agreement with the experimental data within the error bars
except for $\Sn124$.

\begin{figure}[!h]
\begin{center}
\includegraphics[width=0.5\textwidth]{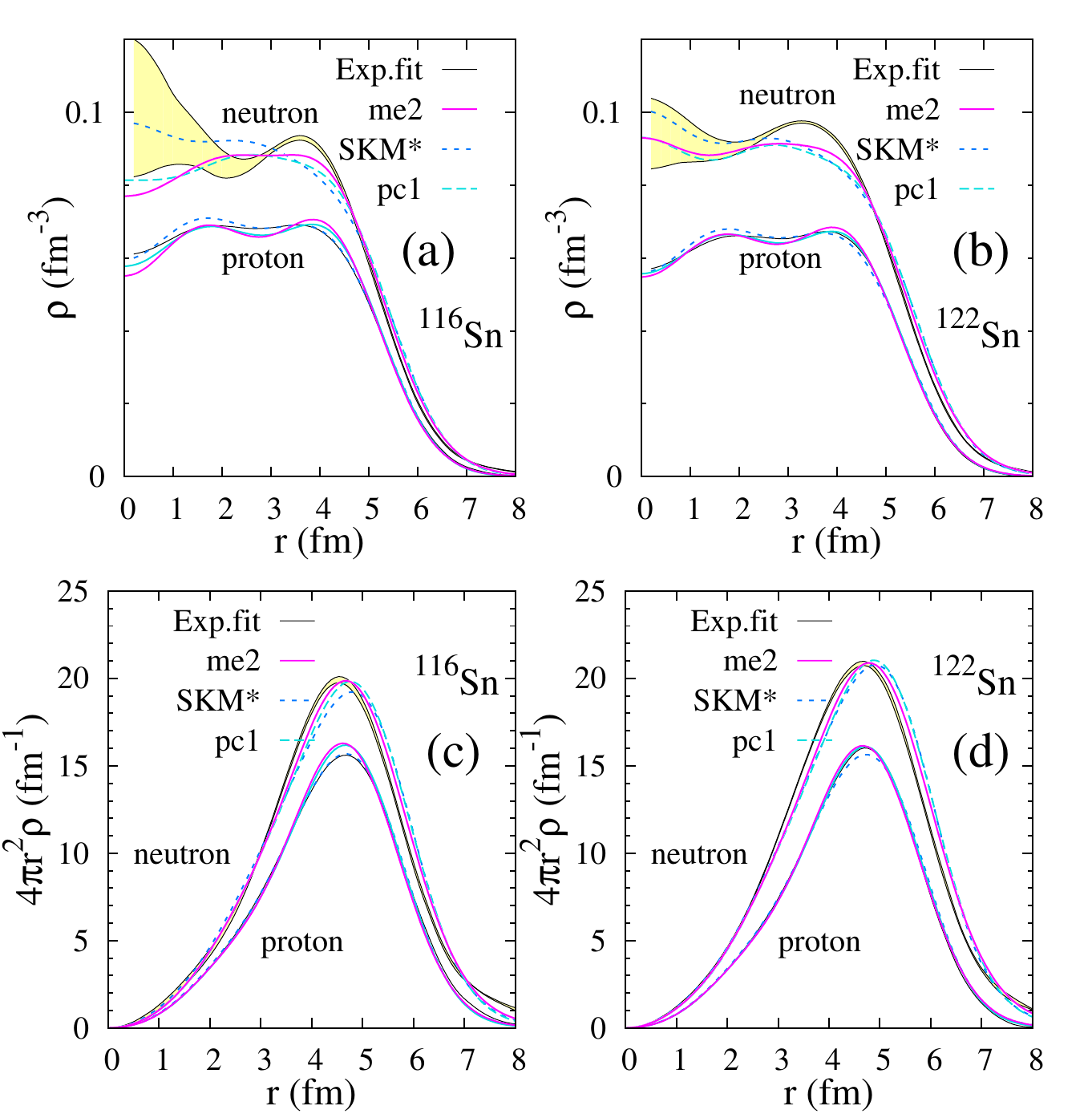}
\end{center}
\caption{Neutron ($\rho^A_{n}$) and proton ($\rho^A_{p}$) density distributions for  $\Sn116$ and $\Sn122$.
(a) and (b) Neutron and proton density distributions for $\Sn116$ and $\Sn122$, respectively,
as obtained from the me2, pc1, and SKM* calculations.  (c) and (d) The corresponding values of
$4\pi r^2\rho$.
The experimental neutron and proton densities from Ref.~\cite{Terashima:2008zza} are also presented (Exp:fit).
The neutron density (error envelopes surrounded by thin lines) was
determined by proton elastic scattering at 295~MeV, while the proton density (thin lines) was obtained from
the charge distribution determined by electron elastic scattering.
\label{fig:density}}
\end{figure}

The theoretical results of the neutron~$(\rho_n)$ and proton~$(\rho_p)$ densities of $\Sn116$ and $\Sn122$ are presented in Fig.~\ref{fig:density}, together with the experimental data from Ref.~\cite{Terashima:2008zza}.
Comparing the me2, pc1, and SKM* results,
all calculations obtain approximately consistent $\rho_p$
and  reproduce the experimental proton density
reasonably well; however, the results for $\rho_n$ depend somewhat on the structure models.
As illustrated in Figs.~\ref{fig:density}~(c) and \ref{fig:density}~(d), the peak position of $4\pi r^2\rho_n(r)$ shifts  slightly to the right in the pc1 and SKM* results
compared with the me2 results.

\subsection{Sn$(p,p)$ cross sections at 295~MeV}

\begin{figure}[!h]
\includegraphics[width=8.5 cm]{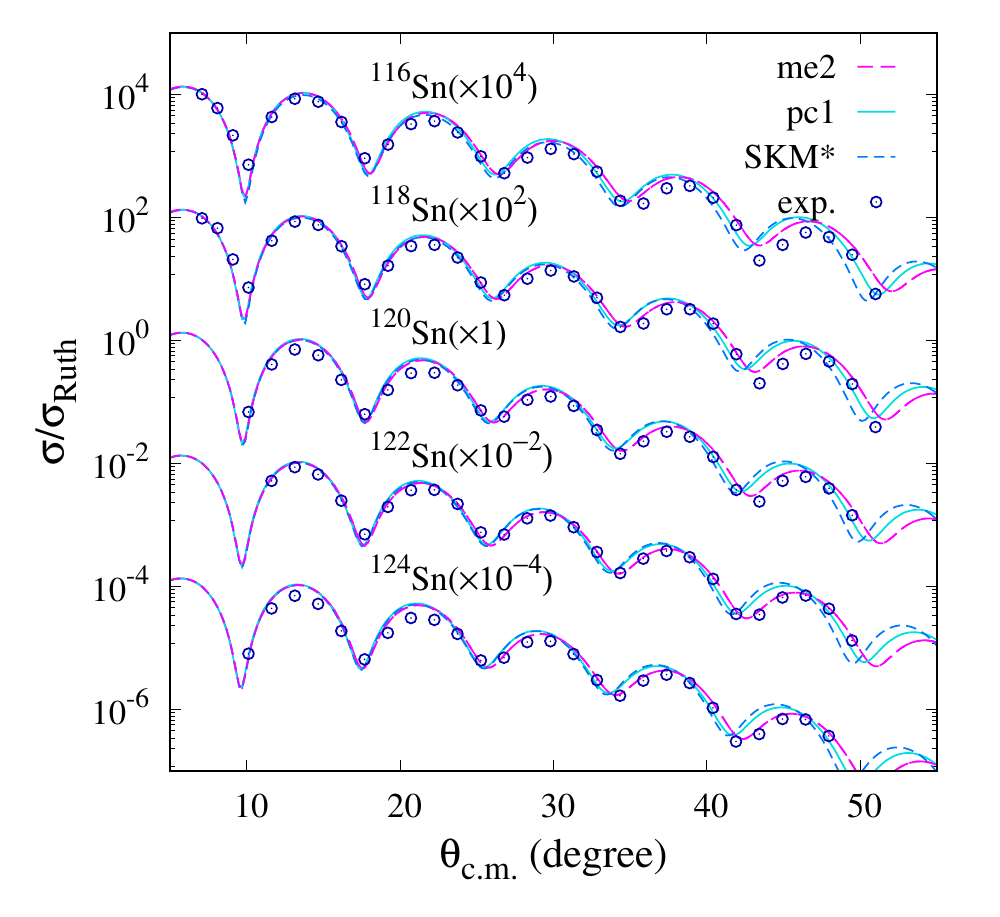}
\caption{Rutherford ratios of the Sn$(p,p)$ cross sections at 295~MeV obtained from the RIA+ddMH calculations
using the me2, pc1, and SKM* densities, together with the
experimental data~\cite{Terashima:2008zza}.
\label{fig:cross-skm}}
\end{figure}

Figure~\ref{fig:cross-skm} presents
the Sn$(p,p)$ cross sections at 295~MeV
obtained from the RIA+ddMH calculations using the theoretical densities,
together with the experimental data.
The calculated cross sections at backward angles are sensitive to the
detailed profile of the surface neutron densities.
Compared with the experimental cross sections,
the me2 density yields cross sections that are in reasonable agreement with the data,
whereas the pc1 and SKM* densities do not.
In the results calculated using the pc1 and SKM* densities,
the diffraction pattern is shrunk, that is, 
the peak and dip positions at backward angles 
shift to forward angles and deviate significantly from the experimental data.
This shrinkage of the diffraction pattern is caused by the outward shift
of the surface-peak position of $4\pi r^2\rho(r)$ in the pc1 and SKM* densities, as illustrated 
previously in Figs.~\ref{fig:density}~(c) and \ref{fig:density}~(d).

The me2 results are compared with the experimental cross sections in detail below.
Calculation with the me2 density
reproduces the $\Sn122(p,p)$ cross sections fairly well.
For the $\Sn116(p,p)$
and $\Sn118(p,p)$ cross sections, calculation with the me2 density successfully reproduces the cross sections
at forward angles; however, the agreement at backward angles is not satisfactory. Namely, 
the calculation somewhat 
overestimates the experimental cross sections.
This overestimation at backward angles can be improved by
a slight modification of the me2 densities for $\Sn116$ and $\Sn118$.

For $\Sn120(p,p)$ and $\Sn124(p,p)$ cross sections, all calculations
overestimate the peak amplitude of the data by 20--30\%
in the entire range of angles. 
In other words, the experimental cross sections for $\Sn120(p,p)$ and $\Sn124(p,p)$ seem
unexpectedly small in the isotopic systematics. 
It is difficult to understand 
this global deviation from the experimental data
because the peak amplitudes at forward angles $\theta_\textrm{c.m.}\sim 13^\circ$
are not sensitive to the detailed neutron density profile.
The normalization of the $\Sn120(p,p)$ and $\Sn124(p,p)$ data should be verified
In the following analysis, I exclude the $\Sn120(p,p)$ and $\Sn124(p,p)$ data
and mainly use the $\Sn116(p,p)$ and $\Sn122(p,p)$ data.

\section{Isotopic analysis}\label{sec:analysis}


Isotopic analysis using the systematics of the densities and cross sections
in isotope chains is a powerful tool to extract structure information from the $(p,p)$ data with less model uncertainty, as reported in a previous study~\cite{Kanada-Enyo:2021oee}.
For the isotopic analysis of $\Sn116(p,p)$ and $\Sn122(p,p)$,
I define the isotopic cross section ratio
of $\Sn122$ to $\Sn116$ as
\begin{align}
R(\sigma;\theta_\textrm{c.m.})\equiv \frac{d\sigma(\Sn122)/d\Omega}{d\sigma(\Sn116)/d\Omega},
\end{align}
where $d\sigma(\SnA)/d\Omega$ is the differential cross sections for the $\SnA(p,p)$ reactions
in the center-of-mass frame.
For the experimental values of $R(\sigma;\theta_\textrm{c.m.})$, I omitted the
mass difference in the transformation from the laboratory to center-of-mass frames.

\begin{figure}[!h]
\includegraphics[width=0.5\textwidth]{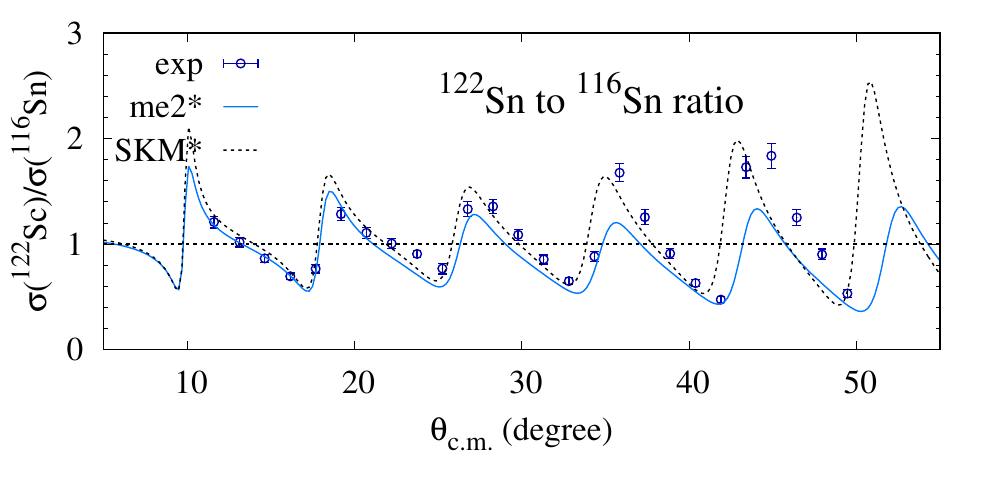}
\caption{ Isotopic cross section ratio $R(\sigma)$ of $\Sn122(p,p)$ to $\Sn116(p,p)$ at 295~MeV. Theoretical results obtained by the
RIA+ddMH calculations using the me2 and SKM* densities are
compared with the experimental data~\cite{Terashima:2008zza}.
\label{fig:cross-compare-skm}}
\end{figure}

The isotopic cross section ratio $R(\sigma)$ at 295~MeV
calculated using the me2 and SKM* densities are displayed in Fig.~\ref{fig:cross-compare-skm},
together with the experimental values.
The ratio $R(\sigma)$ exhibits an oscillatory behavior that corresponds to the slight
difference in the diffraction pattern of the cross sections between $\Sn116$ and $\Sn122$
i.e., the diffraction pattern of $\Sn122$ slightly shrinks due to the increase in $r_n$
compared with $\Sn116$).
In the isotopic analysis, the amplitude of the $R(\sigma)$
oscillation is important because it is
sensitive to the detailed profile of the surface neutron density difference
between $\Sn116$ and $\Sn122$, whereas 
the interval of the oscillation of $R(\sigma)$ is less important.
The results obtained using the me2 density underestimate the
amplitude of the $R(\sigma)$ oscillation of the experimental data, particularly,
at backward angles, whereas
the calculation using the SKM* density well reproduces the
peak magnitude of $R(\sigma)$, although the peak positions deviate somewhat from the data.
The main origin of this difference in the peak magnitude between the two results is
the single-particle occupation of the neutron $3s_{1/2}$ orbit, which significantly affects the surface neutron density.
Table~\ref{tab:occ} lists
the occupation probabilities of
neutron $3s_{1/2}$, $2d_{3/2}$, and $1h_{11/2}$ in $\Sn122$ and $\Sn116$ obtained by
the me2 and SKM* calculations, in comparison with the experimental data determined
by neutron transfer and pick-up experiments~\cite{Szwec:2021uwo}.
For each orbit, the isotopic difference of the occupations 
between $\Sn122$ and $\Sn116$ is also listed.
The isotopic difference of the $3s_{1/2}$ occupation is 
as large as 0.46 
in the me2 results, which is three times larger than the experimental value of 0.15,
whereas  in the SKM* results it is in reasonable agreement with the experimental value.
In other words, the me2 calculation yields 
a rapid increase of the $3s_{1/2}$ occupation from $\Sn116$ to
$\Sn122$, which is inconsistent with the modest change of the 
occupation in the experimental data 
and the SKM results.

\begin{figure}[!h]
\includegraphics[width=8.5 cm]{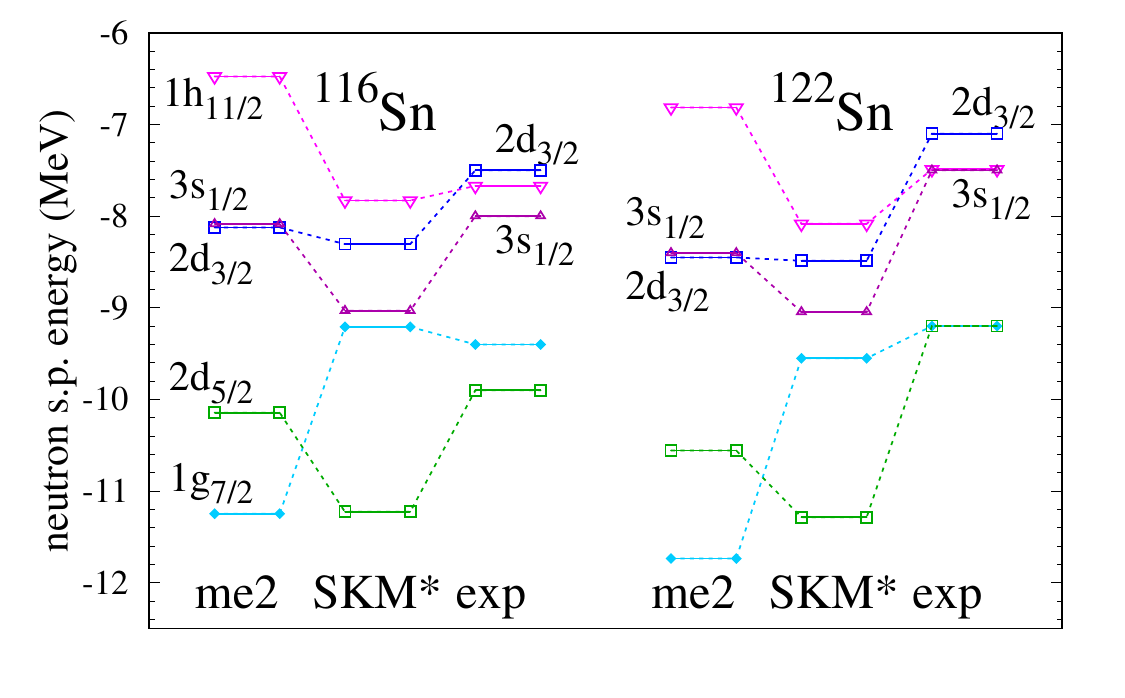}
\caption{Neutron single-particle energies in $\Sn116$ and
$\Sn122$ obtained from the me2 and SKM* calculations
compared with the experimental data from Ref.~~\cite{Szwec:2021uwo}.
\label{fig:spe}}
\end{figure}

\begin{figure}[!htpb]
\includegraphics[width=7 cm]{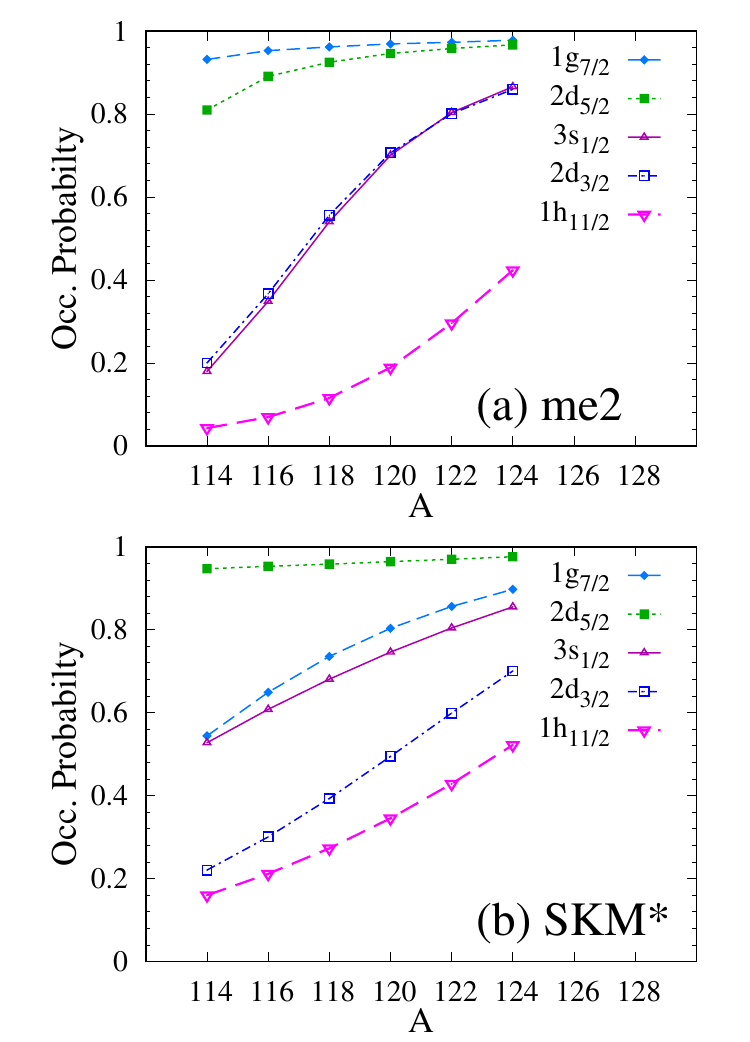}
\caption{Occupation probabilities
of neutron single-particle orbits in the Sn isotopes obtained
from the (a)me2 and (b)SKM* calculations.
\label{fig:occ}}
\end{figure}

The neutron single-particle energies obtained by the me2 and SKM* calculations 
are presented in Fig.~\ref{fig:spe}, together with the experimental data, while
the occupation probability of neutron single-particle orbits in Sn isotopes
is presented in Fig.~\ref{fig:occ}.
In the me2 results, the $3s_{1/2}$ and $2d_{3/2}$ almost degenerate and
they have an approximately 2~MeV energy gap from the lower orbits
($2d_{5/2}$ and $1g_{7/2}$) [see Fig.~\ref{fig:spe}],
As a result, the $3s_{1/2}$ and $2d_{3/2}$ occupation
rapidly increases with an increase in $N$ in the me2 case [Fig.~\ref{fig:occ}(a)]. 
In contrast, the SKM* calculation obtains a lower $3s_{1/2}$ orbit
than the $2d_{3/2}$ orbit.
In this case, the $3s_{1/2}$ orbit is already half-occupied in $\Sn114$, 
and the $3s_{1/2}$ occupation gradually changes
with the increase in $N$ [Fig.~\ref{fig:occ}(b)].
This $3s_{1/2}$-$2d_{3/2}$ level spacing in the SKM* calculation is consistent with 
that of the experimental single-particle energies and yields 
a modest $N$ dependence of the $3s_{1/2}$ occupation, 
contrast to the me2 results.

\begin{table}[!htpb]
\caption{
Occupation probabilities of the $3s_{1/2}$, $2d_{3/2}$, and
$1h_{11/2}$ orbits in $\Sn116$ and $\Sn122$ obtained from the me2 and SKM* calculations, together with the experimental
data~\cite{Szwec:2021uwo}. The isotopic difference in occupations 
between $\Sn122$ and $\Sn116$ is also listed.
\label{tab:occ}
}
\begin{center}
\begin{tabular}{rrrrrccccccccc}
\hline
\hline
&	ME2	&	SKM*	&	exp	\\
\multicolumn{4}{c}{$3s_{1/2}$} \\
$\Sn116$	&	0.35 &	0.61 &	0.42(5)	\\
$\Sn122$	&	0.80 &	0.80 &	0.56(5)	\\
diff. &	{\bf 0.46} &	{\bf 0.20} &	{\bf 0.14(7)}	\\
&\\
\multicolumn{4}{c}{$2d_{3/2}$} \\
$\Sn116$	&	0.37 &	0.30 &	0.15(5)	\\
$\Sn122$	&	0.80 &	0.60 &	0.39(5)	\\
diff. &	{\bf 0.43} &	{\bf 0.30} &	{\bf 0.24(7)}	\\
&\\
\multicolumn{4}{c}{$1h_{11/2}$} \\
$\Sn116$	&	0.07 &	0.21 &	0.29(5)	\\
$\Sn122$	&	0.30 &	0.43 &	0.54(5)	\\
diff.&	{\bf 0.23} &	{\bf 0.22} &	{\bf 0.25(7)}	\\
\hline
\hline
\end{tabular}
\end{center}
\end{table}

\begin{figure}[!htpb]
\begin{center}
\includegraphics[width=7cm]{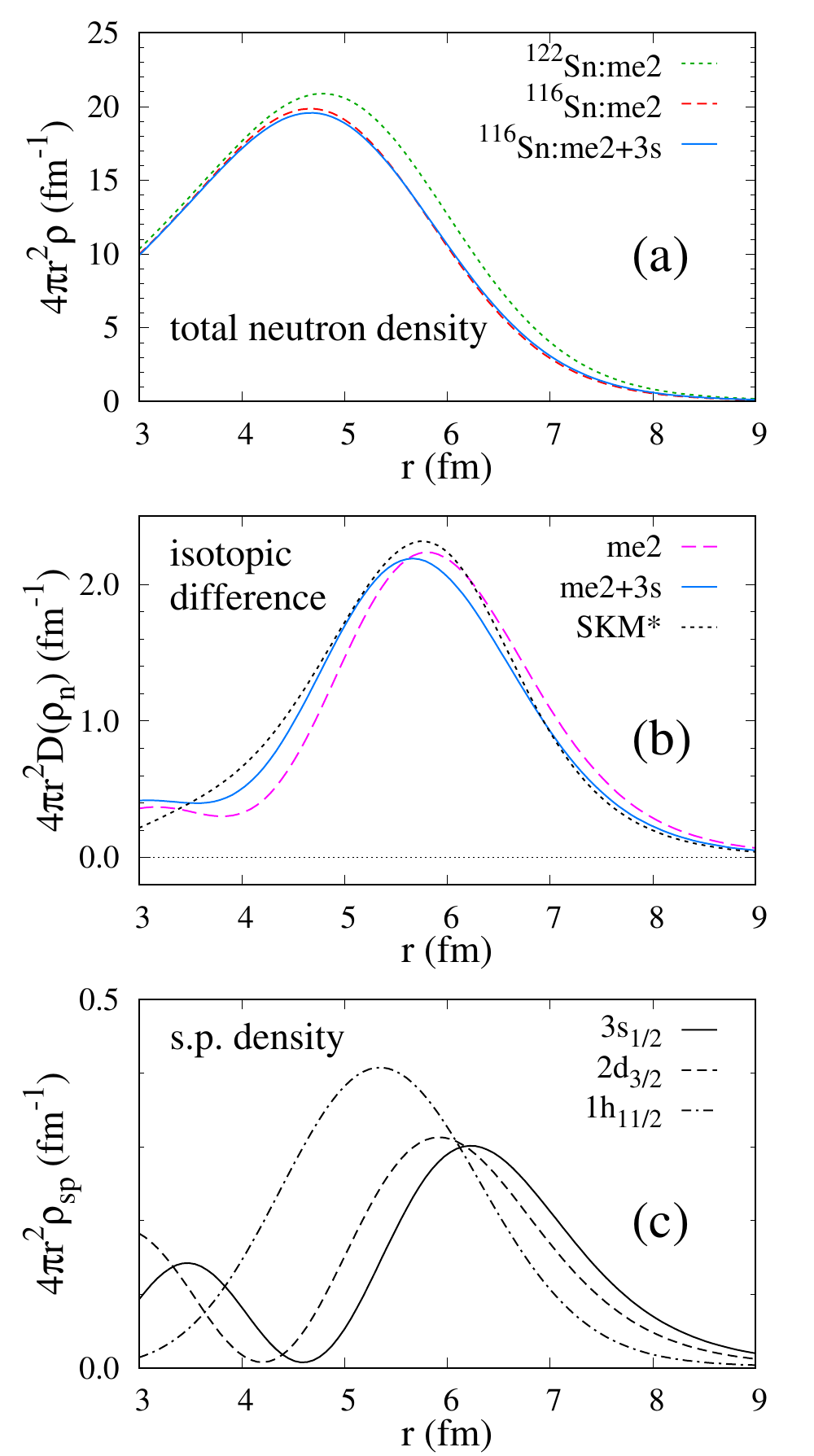}
\end{center}
\caption{(a) Neutron densities of $\Sn116$ for the
me2 and me2+3s densities. For reference, 
the neutron density of $\Sn122$ is also presented.
(b) Isotopic difference in the neutron density of $\Sn122$ from $\Sn116$ obtained for the me2, me2+3s, and SKM* densities.
(c) Single-particle densities $\rho^\textrm{s.p.}(r)$ of the
$3s_{1/2}$, $2d_{3/2}$, and $1h_{11/2}$ orbits in $\Sn120$ obtained from the
me2 calculation.
\label{fig:dens-iso-mod}}
\end{figure}

The failure in reproducing the peak magnitude of $R(\sigma)$ of the $(p,p)$ reaction, presented in Fig.~\ref{fig:cross-compare-skm}, is caused by the rapid
increase of the $3s_{1/2}$ occupation from $\Sn116$ to $\Sn122$
in the me2 calculation throughits contribution to the surface neutron density.
To demonstrate the sensitivity of $R(\sigma)$ to the change of the
$3s_{1/2}$ occupation, I performed a model analysis by changing the
$3s_{1/2}$ occupation in $\Sn116$ from the original me2 result by hand. 
The model density of $\Sn116$ is given as
\begin{align}
\rho^{116}_{n,\textrm{model}}(r)&=(1-\frac{1}{N})\rho^{116}_n(r)+\rho^\textrm{s.p.}_{3s_{1/2}}(r),
\end{align}
where $\rho^\textrm{s.p.}_{3s_{1/2}}(r)$ is the single-particle density of
the neutron $3s_{1/2}$ orbit.
In this model (called the ``me2+3s model'') the density in $\Sn122$ was unchanged, while
the $3s_{1/2}$ occupation number in $\Sn116$ was increased by 1 
from the original me2 density, corresponding to the increase of 
the $3s_{1/2}$ occupation probability by 0.5. 
After the increase, 
the $3s_{1/2}$ occupation in $\Sn116$ became almost the same as that in $\Sn122$.
Figure~\ref{fig:dens-iso-mod}(a) compares the me2 and me2+3s neutron densities of $\Sn116$, 
together with those in $\Sn120$, while
Fig.~\ref{fig:dens-iso-mod}(c)
represents the neutron single-particle densities of  $3s_{1/2}$, $2d_{3/2}$, and
$1h_{11/2}$ obtained by the me2 calculation for $\Sn120$.

Due to the modification of the $3s_{1/2}$ occupation in $\Sn116$, 
the surface neutron density
around $r=4$--$5$~fm
decreases from the original me2 density 
[see blue solid and magenta dashed lines at the peak of $4\pi r^2$ in Fig.~\ref{fig:dens-iso-mod}(a)].
To examine the isotopic change of the neutron density more closely, I defined
the isotopic neutron density difference
between $\Sn116$ and $\Sn122$ as
\begin{align}
&D(\rho_{n};r)\equiv \rho_{n}(\Sn122;r)-\rho_{n}(\Sn116;r).
\end{align}
Fig.~\ref{fig:dens-iso-mod}(b) represents the values of $4\pi r^2 D(\rho_{n};r)$ obtained for 
the me2, me2+3s, and SKM* densities.
$D(\rho_{n};r)$ of the me2 density is dominantly contributed by the major-shell
orbits, $3s_{1/2}$, $2d_{3/2}$, and $1h_{11/2}$.
$D(\rho_{n};r)$ around $r=4$--$5$~fm is significantly larger 
in the me2+3s and SKM* densities than the me2 density
due to the modest change of the $3s_{1/2}$ occupation 
from $\Sn116$ to $\Sn122$. In contrast, 
$D(\rho_{n};r)$ at the nuclear surface is smaller because 
the rapid increase of the $3s_{1/2}$ occupation reduce the 
$D(\rho_{n};r)$ in the  $r=4$--$5$~fm region. 

\begin{figure}[!htpb]
\includegraphics[width=0.5\textwidth]{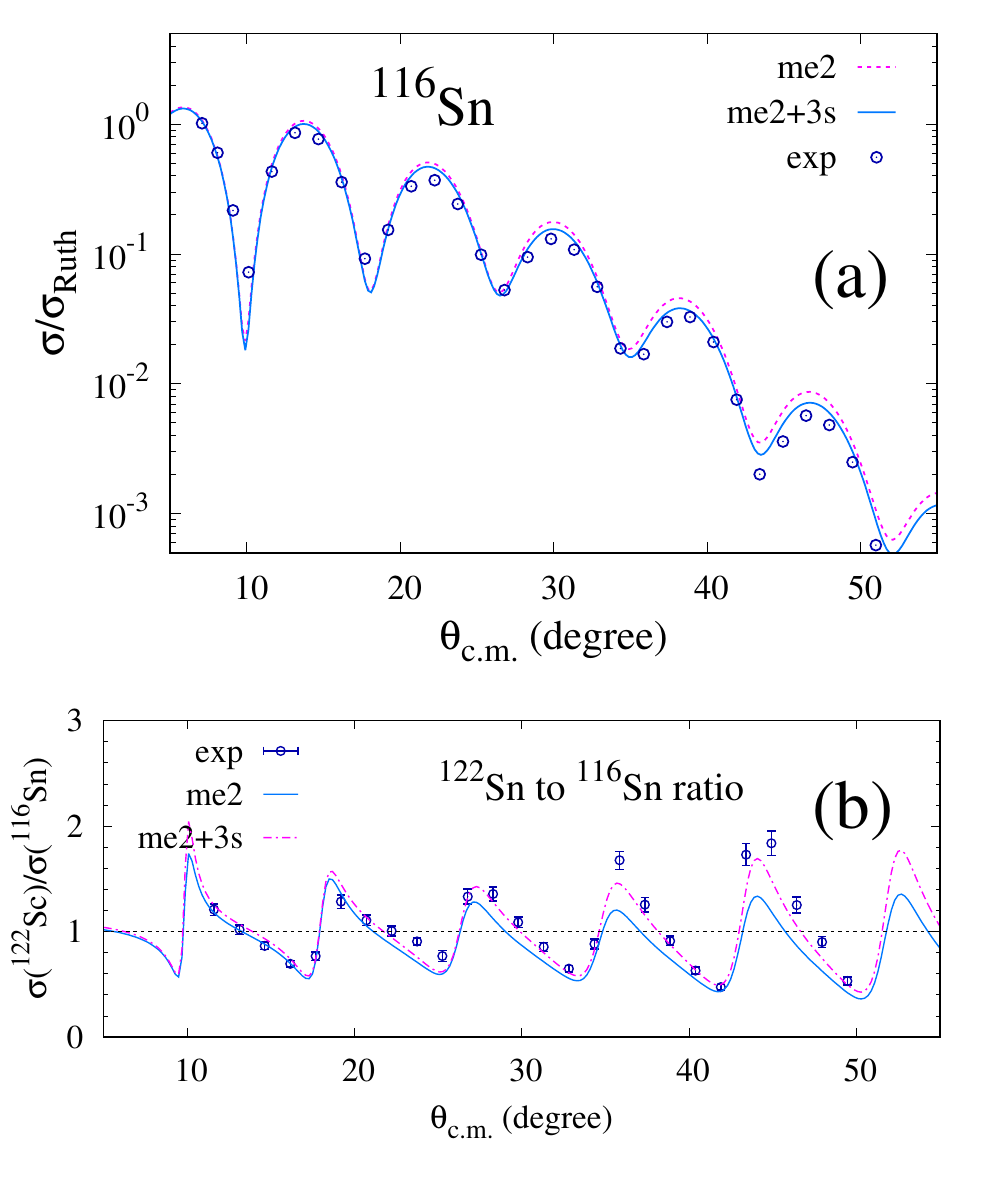}
\caption{
(a) The $\Sn116(p,p)$ cross sections at 295~MeV
and (b) isotopic cross section ratios $R(\sigma)$ of $\Sn122(p,p)$ to $\Sn116(p,p)$ obtained by the
RIA+ddMH calculations using the me2 and me2+3s densities, compared with
the experimental data~\cite{Terashima:2008zza}.
\label{fig:cross-compare-mod}}
\end{figure}

Using the modified $\Sn116$ density of me2+3s, I calculated the
$\Sn116(p,p)$ cross sections and the isotopic cross section ratio
$R(\sigma)$. Figure~\ref{fig:cross-compare-mod} compares
the results obtained by employing the me2+3s and me2 densities.
As expected, the me2+3s density yields an improved result of the
$\Sn116(p,p)$ cross sections at backward angles [Fig.~\ref{fig:dens-iso-mod}(a)].
Moreover, the result of $R(\sigma)$
is significantly improved in the me2+3s calculation
and agrees well with the experimental data [Fig.~\ref{fig:dens-iso-mod}(b)].

In general, $R(\sigma)$ of 295~MeV $(p,p)$ reactions
is a sensitive probe of the occupation change of low-$\ell$ orbits, as discussed 
in previous work~\cite{Kanada-Enyo:2021oee}.
I performed a similar analysis by changing the neutron occupation numbers of 
$2d_{3/2}$ and $1h_{11/2}$ in $\Sn116$ and verified that the peak amplitudes of $R(\sigma)$
were not as sensitive to the $2d_{3/2}$ and $1h_{11/2}$ occupations
as to the $3s_{1/2}$ occupation.
It should be emphasized again that the peak amplitude
of $R(\sigma)$ of $\Sn122$ to $\Sn116$
at backward angles is sensitive to the isotopic neutron density difference 
$D(\rho_{n};r)$ at the nuclear surface around $r=4$--$5$~fm and is
affected by the isotopic change of the $3s_{1/2}$ occupation. 
The conclusion derived from the present analysis is that
a rapid increase of the $3s_{1/2}$ occupation from $\Sn116$ to $\Sn122$
obtained in the me2 calculation is unlikely.

\section{Summary}\label{sec:summary}

The isotopic analysis of Sn isotopes in the range $A=116$--124 was performed
by combining the nuclear structure and reaction calculations.
$\SnA(p,p)$ reactions at 295~MeV were calculated using the RIA+ddMH model
with theoretical densities for the Sn isotopes obtained from RHB and SHFB calculations.
The RIA+ddMH calculations
using the theoretical density from the RHB calculations
with the me2 interaction (i.e., DD-ME2 interaction) were in reasonable agreement
with the experimental cross sections; however, 
they somewhat  overestimated the backward cross sections
of the $\Sn116(p,p)$ and $\Sn118(p,p)$ reactions.

A detailed analysis of the isotopic systematics was performed by using
the isotopic cross section ratios
$R(\sigma)$ of $\Sn122$ to $\Sn116$.
Comparing the theoretical $R(\sigma)$ with the experimental values,
the peak amplitudes of $R(\sigma)$ at backward angles were significantly
underestimated by the me2 density, whereas they were well reproduced by the
SKM* density.
The main cause of the underestimation of the me2 result was found to be
the $3s_{1/2}$ neutron contribution.
The $3s_{1/2}$ occupation probability rapidly
increased with an increase in $N$
in the me2 calculation; however,  this was inconsistent
with the experimental observation of only a small increase
from $\Sn116$ to $\Sn122$.

To demonstrate the sensitivity of $R(\sigma)$ to the $3s_{1/2}$ neutron contribution,
model analysis (me2+3s) was performed
by modifying the neutron $3s_{1/2}$ occupation in $\Sn116$.
It was proved that the $\Sn116(p,p)$ cross sections at backward angles
were sensitive to the $3s_{1/2}$ neutron contribution. This signifies that
the peak amplitudes of $R(\sigma)$ of $\Sn122$ to $\Sn116$
serve as a sensitive probe of the isotopic
change of the neutron $3s_{1/2}$ occupation through its effect to the 
isotopic difference
$D(\rho_{n};r)$ of the surface neutron density in the $r=4$--$5$~fm region.
In the model calculation using the me2+3s density,
the peak amplitudes of $R(\sigma)$ at backward angles were reproduced well.
The good reproduction of the experimental $R(\sigma)$ indicates 
that a rapid increase of the $3s_{1/2}$ occupation
from $\Sn116$ to $\Sn122$ obtained by the me2 calculation is
unlikely. This result derived from the 295 MeV $\SnA(p,p)$ cross section data
\cite{Terashima:2008zza} is consistent with the direct measurement of
neutron occupations in Sn isotopes by the neutron transfer and pick-up
reactions~\cite{Szwec:2021uwo}.

\begin{acknowledgments}
This work was supported
by Grants-in-Aid of the Japan Society for the Promotion of Science (Grant Nos. JP18K03617, JP18H05407, and JP22K03633)
and by a grant of the joint research project of the Research Center for Nuclear Physics at Osaka
University.
\end{acknowledgments}

\bibliographystyle{apsrev4-1}
\bibliography{RIA-pb-refs}

\end{document}